# Single-Carrier Spatial Modulation for the Internet of Things: Design and Performance Evaluation by Using Real Compact and Reconfigurable Antennas

D.-T. Phan-Huy[1], Member, IEEE, Y. Kokar[2], K. Rachedi[3], P. Pajusco[4], A. Mokh[2], T. Magounaki[1], R. Masood[4], C. Buey[1], P. Ratajczak[1], N. Malhouroux-Gaffet[1], J.-M. Conrat[1], J.-C. Prévotet[2], A. Ourir[3], J. de Rosny[3], M. Crussière[2], M. Hélard[2], Member, IEEE, A. Gati[1], T. Sarrebourse[1], M. Di Renzo[5], senior Member, IEEE

[1]Orange Labs Networks, Châtillon, 92326 France
[2]INSA Rennes, CNRS, Institut d'Electronique et de Telecommunication de Rennes, UMR 6164, University of Rennes 1, 35000 Rennes, France
[3]Institut Langevin and Ecole Supérieure de Physique et de Chimie Industrielle, CNRS, 75005 Paris, France
[4]Institut Mines Telecom Atlantique, 29238, Brest, France
[5]Laboratoire des Signaux et Systémes, CNRS, University of Paris-Saclay, CentraleSupélec, University of Paris-Sud, 91192 Gif-sur-Yvette, France

Corresponding author: D.-T. Phan-Huy (dinhthuy.phanhuy@orange.com)

This work was supported in part by SpatialModulation project under the French Research Agency Grant ANR-15-CE25-0016.

**ABSTRACT** In this paper, for the first time, we propose two new solutions to boost the data rate between small connected objects such as glasses and cams and the 5th generation (5G) mobile network, based on spatial modulation, single carrier waveform, compact reconfigurable antennas at the object side and massive multiple input multiple output (M-MIMO) at the network side. In the first new wireless communication system, a "transmitting object" uses transmit spatial modulation with a compact reconfigurable antenna and a constant envelop amplifier to transmit in high data rate with a low complexity and low power consumption. The space-time digital processing capability of the M-MIMO 5G base station is used to detect such signal. In the second new wireless communication system, a "receiving object" uses receive spatial modulation, a compact multiport antenna and a low complexity detection algorithm to receive in high data rate with a low complexity signal processing. The space-time beamforming capability of the M-MIMO 5G base stations is exploited to deliver a signal that is pre-equalized enough to be detected by the object. For the first time, we present experiments showing that M-MIMO allows for the re-introduction of single carrier modulation waveform. For the first time, we present performance results obtained with real existing compact antennas and compact reconfigurable antennas, showing that the two new communication systems outperform conventional modulation in terms of energy efficiency and complexity.

## I. INTRODUCTION

Future mobile networks of the 5th generation (5G) will provide a wireless connection to the Internet-of-Things (IoT) [1]. Among connected things, some, like connected glasses, connected cameras and connected watches, will need to transmit or receive video streams at a high data rate. The 3rd Generation Partnership Project (3GPP) has already started to lower down the cost and the power consumption of devices for connected objects by reducing the number of Radio Frequency (RF) chains (for transmission) and reducing the number of RF amplifiers [2].

Recently, it has been shown that, for some high signal to noise ratios values, and still with a single RF chain at the transmitter side, by using transmit spatial modulation [3] with conventional arrays of antenna elements [4], one can achieve a higher spectral efficiency than by using a conventional modulation with the same single RF chain [5]. In transmit spatial modulation systems, in addition to the data stream sent using a conventional Pulse shape Amplitude Modulation (PAM), an additional data stream is sent by switching the transmit antenna element, every symbol period. The index of the current transmit antenna element encodes binary information. Previous studies on transmit spatial modulation are focused on conventional arrays: [6] analyzes solutions based on spatial modulation that exploit the diversity gain of the conventional antenna array and [7]-[14] provide preliminary results on the performance of transmit spatial modulation based on experimental data obtained with conventional antenna arrays. For more information on transmit spatial modulation, please refer to the survey papers [15]-[19].

In receive spatial modulation systems such as the ones studied in [20], in addition to sending a conventional PAM, the transmitter sends an additional data stream by focusing towards one antenna of the receiver among several, every symbol period. The index of the target antenna encodes binary information. A simple receiver can be used to demodulate the two streams [21]-[23].

Other works on reconfigurable antennas [24]-[26] suggest that one can transport more bits than with a SISO system by using a reconfigurable antenna. In this case, instead of switching between antenna elements, the transmitter switches between radiation patterns. The advantage of these solutions is that they are more compact in size. The limitation of these techniques is the non-orthogonality of the radiation patterns, and up to now, no complete performance study with real and existing compact antennas has been performed to verify the advantage of such antennas.

In parallel, 5G envisions the development of massive arrays [28]-[30] at the network side, with hundreds of antenna elements. These Massive Multiple Input Multiple Output (M-MIMO) antennas benefit from higher beamforming and spatial multiplexing gains [28][29], or reduced complexity in the demodulation [30].

In this paper, for the first time, we introduce and make a complete performance evaluation of two new communication systems illustrated in Figure 1, for uplink and downlink communications, respectively, between objects with compact antennas and a base station with M-MIMO antenna array. In the uplink, transmit spatial modulation is used, and the object antenna is a compact reconfigurable antenna (either a multiport switchable antenna or a monoport reconfigurable antenna), whereas in the downlink, receive spatial modulation is used and the object antenna is still a compact antenna but multiport only in the current implementation. These two systems are compared to systems using conventional modulation schemes as well.

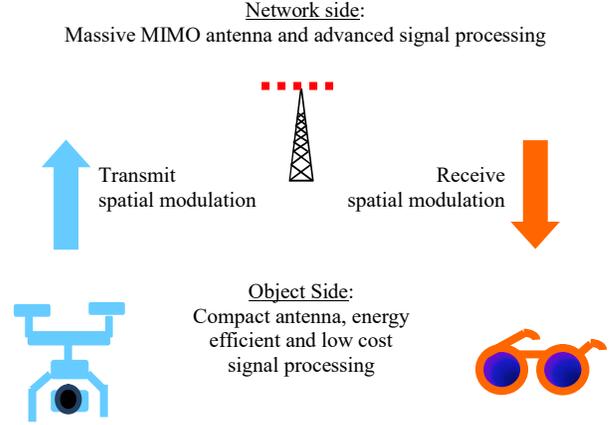

FIGURE 1. Transmit and receive spatial modulation for connected objects

The paper is organized as follows: Section II recalls the transmit and receive spatial modulation concepts and illustrates these concepts with visual experiments. Section III and IV present the performance evaluation study for transmit spatial modulation and receive spatial modulation, respectively.

The following notation is used throughout the paper. If $x \in \mathbb{C}$, then $|x|$ is its module, $\arg(x)$ is its phase in radians, $x^*$ is its conjugate, $\mathcal{R}e(x)$ is its real part and $\text{Im}(x)$ is its imaginary part. If $\mathbf{H} \in \mathbb{C}^{M \times P}$, then $\mathbf{H}^\dagger$ is the transpose-conjugate of $\mathbf{H}$ and $\|\mathbf{H}\|^2 = \sum_{p=0}^{P-1} \sum_{m=0}^{M-1} |\mathbf{H}_{m,p}|^2$. If $\mathbf{x} \in \mathbb{C}^{P \times 1}$, then $\|\mathbf{x}\|^2 = \sum_{p=0}^{P-1} |\mathbf{x}_p|^2$. $[n, p]$ is the set of integers between $n$ and $p$, including $n$ and $p$. $j^2 = -1$. We define the set $\mathcal{B}^{(P)}$, $S^{\text{QPSK}}$, $S^{\text{8PSK}}$ and $S^{\text{16QAM}}$ as follows:

- $\mathcal{B}^{(P)} = \{\mathbf{b}^{(l)} \in \{0,1\}^{K \times 1} | l \in [1, P]; K = \log_2(P); \sum_{k=1}^{K} \mathbf{b}_k^{(l)} 2^{k-1} = l\}$;
- $S^{QPSK} = \{\frac{1+j}{\sqrt{2}}, \frac{1-j}{\sqrt{2}}, \frac{-1-j}{\sqrt{2}}, \frac{-1+j}{\sqrt{2}}\}$;
- $S^{8PSK} = \{e^{\frac{2j\pi(n-1)}{8}}, n \in [\![1; 8]\!]\}$;
- $S^{16QAM} = \{\frac{-3+2l+(-3+2k)j}{\sqrt{10}} | l \in [0,3] \text{ and } k \in [0,3]\}$.

## II. TRANSMIT AND RECEIVE SPATIAL MODULATION CONCEPTS

### A. TRANSMIT SPATIAL MODULATION

In this section, we recall the concept of transmit spatial modulation and illustrate this concept with a visual experiment illustrated in Figure 2.

During the communication, the transmitter activates one radiation pattern among $P = 2^K$ distinct ones. Each radiation pattern is associated with a distinct binary sequence of $K$ bits, according to a pre-defined pattern-to-bit mapping rule. The rule is known at both the transmitter and the receiver sides. As in most wireless communication systems, the transmitter sends pilots to train the receiver. More precisely, the transmitter activates each of its radiation patterns alternatively, so that the receiver can estimate and store the propagation channel associated with each pattern. Then, the transmitter performs an actual data transmission as follows. A long binary sequence is cut into sub-sequences of $K$ bits. To transmit a particular sub-sequence, the transmitter activates the pattern that corresponds to the sub-sequence, using the pattern-to-bit mapping rule. The receiver detects the pattern that has been used by comparing the current received channel with the $P$ stored channels. The receiver converts the detected pattern into a sub-sequence of $K$ bits using the pattern-to-bit mapping rule. Note that, transmit spatial modulation can be combined with a conventional modulation.

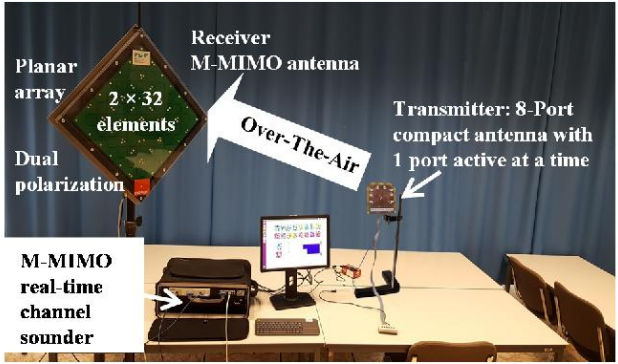

**FIGURE 2.** Experimental set-up

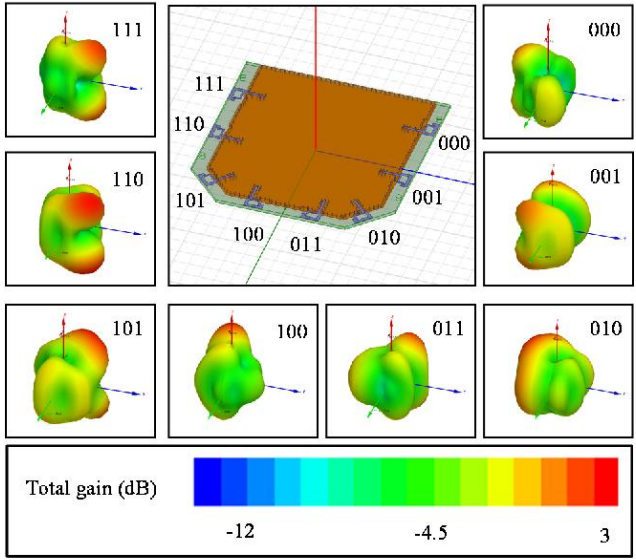

**FIGURE 3.** The 8 ports of the transmitter, their corresponding radiation patterns and their corresponding 3-bit sequences

As illustrated in Figure 2, our visual experiment involves an 8-port compact antenna at the transmission side and a 64-element M-MIMO antenna at the receiver side. The transmitter emits a signal with carrier frequency of 2.45 GHz with one antenna port among the $P = 8$ available, thanks to a switch. In this case, a pattern encodes $K = 3$ bits. As illustrated in Figure 3, the 8 radiation patterns corresponding to the 8 ports of this antenna are distinct. Figure 3 also provides the mapping rule between patterns and 3-bit sequences.

As illustrated in Figure 2, the 64 elements of the M-MIMO antenna are positioned on a grid of 8 lines and 8 columns and connected to a real-time channel sounder. Figure 4 illustrates the graphical interface of the sounder. The top of Figure 4 illustrates the result of a previous channel training phase. For each of the 8 radiation patterns, the spatial signature (i.e., the matrix of 8 by 8 estimated channel amplitudes) is stored and displayed in color scale. Each spatial signature is associated with a pattern and a 3-bit sequence. The bottom of Figure 4 illustrates the data transmission. A 3-bit sequence is transmitted by activating one radiation pattern among the 8 available. The channel sounder displays the current spatial signature (i.e., the actual 8 by 8 matrix of amplitudes of the current channel) in a color scale, computes the correlation of this matrix with the 8 stored matrices, determines the pattern that maximizes the correlation, and converts it into the detected 3-bit sequence, using the pattern-to-bit mapping rule. As illustrated in a video [32], all these steps are done in real time. In [32], even though the channel training is done only once at the beginning of the video, the detection remains robust for a long time period. This is thanks to the large size of the 8 by 8 matrices on which the correlation is performed. This illustrates one of the advantages of using M-MIMO at the network side.

Note that the training phase of spatial modulation is equivalent to the channel estimation phase for a conventional single carrier modulation. Both modulations require the transmission of pilot symbols during these phases, with the same periodicity (enough to track the channel variation).

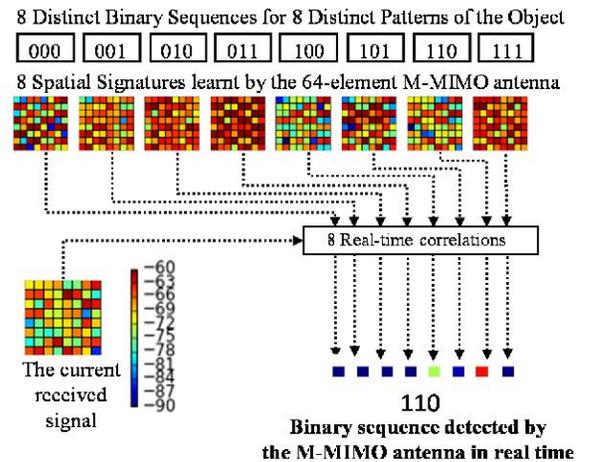

**FIGURE 4.** Graphical interface of the M-MIMO real time channel sounder

### B. RECEIVE SPATIAL MODULATION
In this section, we recall the concept of receive spatial modulation and illustrate this concept with a visual experiment depicted in Figure 5.

The receiver has $P = 2^K$ antenna ports with $P$ distinct associated radiation patterns. Each radiation pattern is associated with a distinct binary sequence of $K$ bits, according to a pre-defined pattern-to-bit mapping rule. The rule is known at both the transmitter and the receiver sides. Channel reciprocity based beamforming (BF) as implemented in time division duplex (TDD) mode systems is used in order to target one antenna of the receiver among $P$. As in any channel reciprocity based BF system, during the training phase the receiver sends distinct pilots from its distinct ports so that, for each port, the transmitter can: 1) estimate the channel, and 2) compute and store the precoder that enables to beamform towards this particular port. Then, the transmitter performs data transmission as follows. A long binary sequence is cut into sub-sequences of $K$ bits. To transmit a particular sub-sequence, the transmitter uses the stored precoder that beamforms towards the pattern corresponding to the considered sub-sequence. Finally, after detecting the port that is the current target of the beamforming (for instance by identifying the port that receives the strongest power), the receiver converts the detected pattern into a sub-sequence of $K$ bits using the pattern-to-bit mapping rule. Note that, this can be combined with a conventional modulation.

In our visual experiment, a transceiver and a receiver detailed in [33] are used with a carrier frequency of 2.48 GHz.

As illustrated in Figure 5, the transmitter is a uniform linear array (ULA) of 4 monopoles and the receiver is a ULA of $P = 4$ "squeezed monopoles". These monopoles are "squeezed" in the sense that they are close to each other by much less than half a wavelength and subject to coupling. In this experiment, the transmitter uses maximum ratio transmission (MRT) beamforming [34], and switches at a maximum speed of 125 kHz between the four different stored precoders, each precoder targeting a distinct "squeezed monopole". In this example, the identity of the target "squeezed monopole" encodes $K = \log_2(P) = 2$ bits. In parallel, the system transmits BPSK symbols.

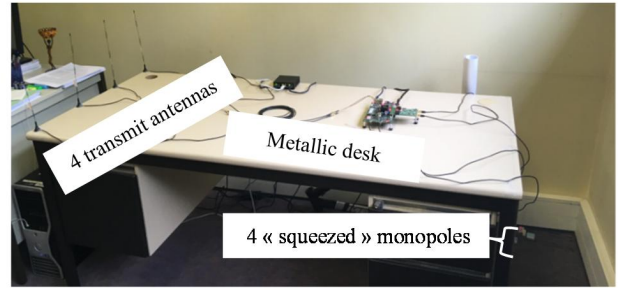

a) Experimental set-up, non line-of-sight (NLOS)

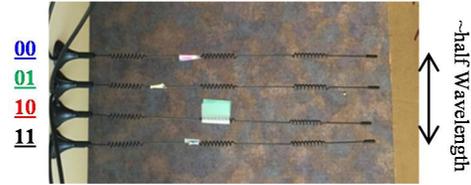

b) Zoom on the 4 « squeezed » monopoles

**FIGURE 5.** Experimental set-up.

Figure 6 illustrates the resulting constellation of receive spatial modulation together with BPSK modulation. The positions of the two BPSK states in the complex I-Q domain are visualized by two drawings of targets. The I-Q symbol received by each of the $P = 4$ monopoles of the receiver is plot for different realizations of the propagation channel. The joint detection of BPSK and the spatial modulation consists in determining the monopole that is closest to one of the two BSPK states. Figure 6 illustrates experimental measurements of the received I-Q symbols. The measurements are classified into 8 different categories illustrated by 8 different figures. For each category, the same BPSK symbol and the same spatial modulation symbol are detected. One can note that spatial modulation works even with these "squeezed monopoles". This is due to the fact that for one monopole the other monopoles act as parasitic scatterers. This is known to create decorrelation, and is exploited as a design principle for compact multiport antennas as the ones that were presented in [35][36], and that will be used in the current paper for performance evaluation.

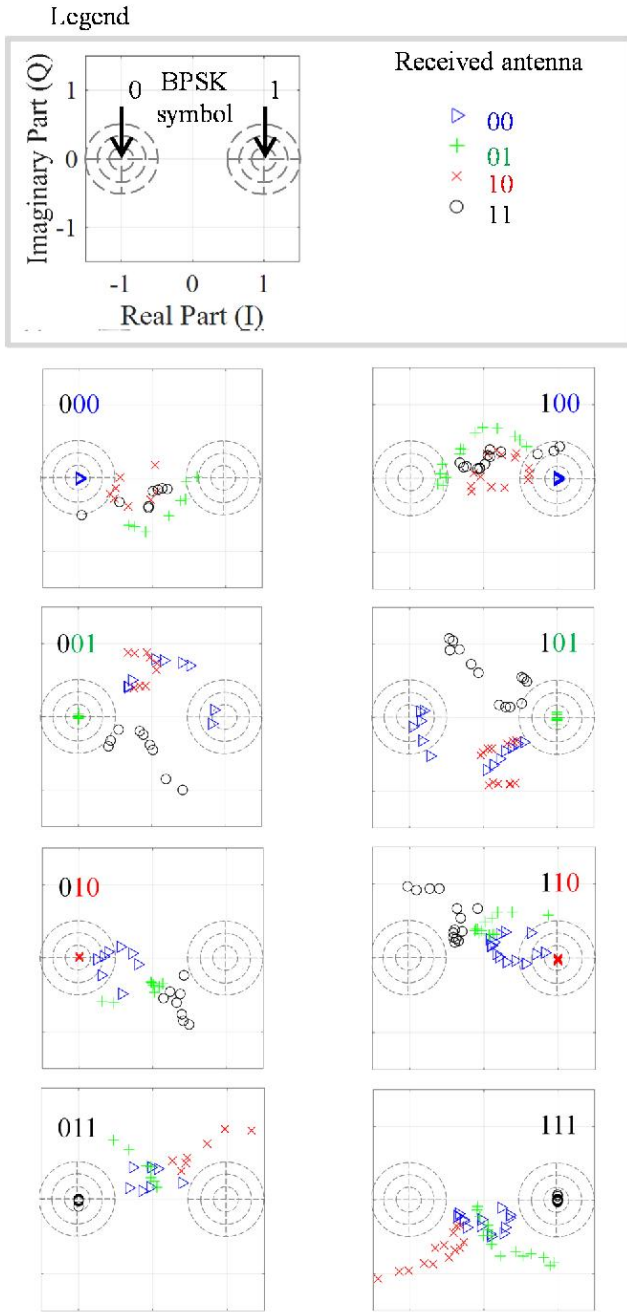

**FIGURE 6.** Experimental observations of the constellations for receive spatial modulation and BPSK modulation.

## III. MASSIVE MIMO: AN ENABLER FOR SINGLE-CARRIER MODULATIONS

5G is based on multi-carrier modulation. Such modulation is not compatible with spatial modulation [17]. However, we believe that the introduction of M-MIMO antennas in 5G networks is an enabler for the re-introduction of single-carrier modulation in the future. In previous works [37][38], it has been shown that thanks to time reversal focusing with a large number of transmit antennas, the signal at the receiver is nearly echo-free. In [38], a single-tap receiver successfully demodulates a signal using a single carrier modulation, within a 30MHz bandwidth centered at 1 GHz and using a 256 Quadrature Amplitude Modulation (QAM). However, this experiment was performed using an arbitrary waveform generator at the transmission side, and an oscilloscope at the reception side.

In this paper, we show recent experimental results obtained using the open-source hardware and software development platform Open Air Interface (OAI), and a rail moving with a Digital Servo Amplifier, SERVOSTAR 300, along with a Rosier servo motor controlling the movement. OAI is a wireless technology platform that offers an open-source software-based implementation of the Long Term Evolution (LTE) system spanning the full protocol stack of 3rd Generation Partnership Project (3GPP) standard both in Evolved Universal Terrestrial Radio Access Network (E-UTRAN) and Evolved Packet Core (EPC). The experiments were carried out using orthogonal frequency division multiplex (OFDM) frames at the carrier frequency of 2.68 GHz. Each OFDM symbol consists of 512 carriers, out of which 300 are filled with random QPSK symbols and the rest are set to zero. An extended cyclic prefix (ECP) of 128 samples is added to each OFDM symbol after the 512-point Inverse Fast Fourier Transform (IFFT). The sampling rate is 7.68 mega symbols per second, resulting in an effective bandwidth of 4.5 MHz. Ten subframes each with 12 ECP-OFDM symbols compose the TDD OFDM frame.

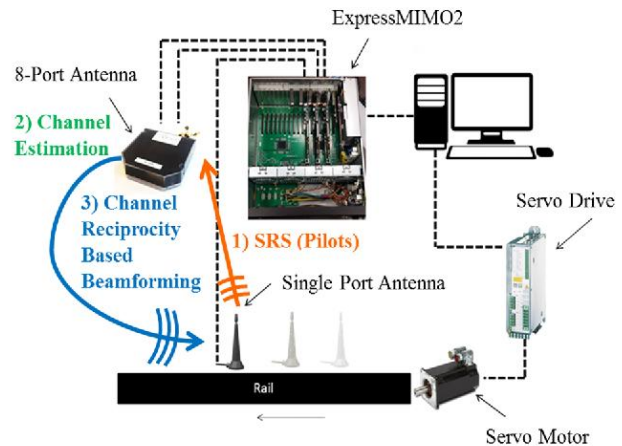

**FIGURE 7.** Experimental set-up, in NLOS.

In the experiment, the receiver sends pilots called sounding reference signals (SRS) in the uplink direction. The transmitter uses these pilots to estimate the uplink channel. Channel reciprocity is exploited to deduce the downlink channel. The transmitter precodes its downlink data and pilots with a maximum ratio transmission (MRT) precoder to beamform the signal towards the receiver. Note that MRT is equivalent to time reversal or the transmit matched filter pre-filtering applied to OFDM instead of a single-carrier modulation [33]. During the experiment, 15 different positions of the receiver are tested, along a rail, and all in NLOS of the transmitter. For each position, the receiver measures the frequency response of the received beamformed channel thanks to the downlink precoded pilots. The measurements were carried out inside a controlled laboratory environment. Figure 7 illustrates the measurement setup.

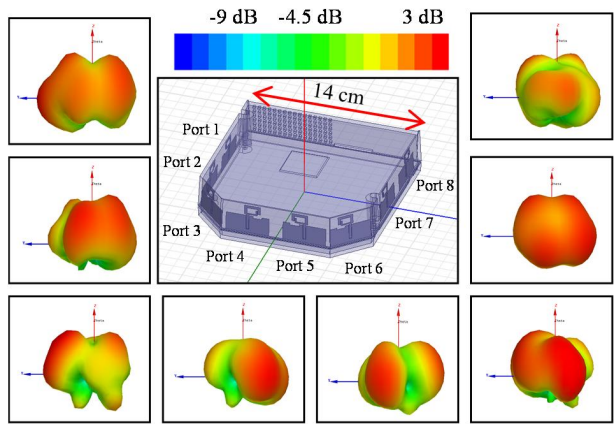

**FIGURE 8.** 8-port transmit antenna.

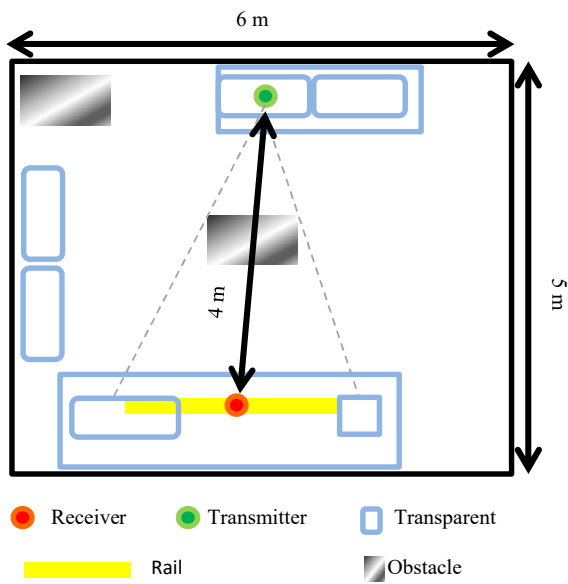

**FIGURE 9.** NLOS propagation.

Figure 8 illustrates the radiation patterns of the 8 ports of the transmitter. Although, the antenna is compact, it exhibits patterns that are diverse. This antenna is therefore equivalent to an array of 8 low-correlated antennas.

Figure 9 illustrates the propagation environment during the experiments. Non-line-of sight propagation is chosen to create multi-path propagation. In such environment, the channel impulse response has several delayed taps. Hence, a single tap receiver trying to demodulate a single carrier modulation would suffer from inter-symbol interference. Finally, we apply an IFFT to the frequency response of the received beamformed channel, to obtain the corresponding filter in the time domain.

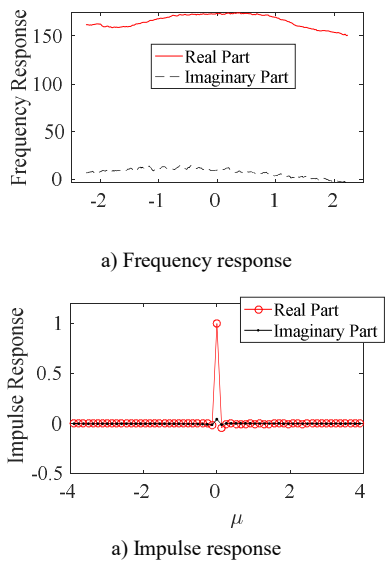

a) Frequency response

a) Impulse response

**FIGURE 10.** Received beamformed channel for the 15th position of the UE.

Figure 10 illustrates the measured frequency response and the corresponding impulse response, for the received beamformed channel measured at the 15$^{th}$ position. We observe that the beamformed channel is nearly a single tap channel. We also evaluate the ratio between the useful signal and the inter-symbol interference (SIR) that would be undergone by a single tap-receiver demodulating a single carrier modulation at 5 MHz. As illustrated in Figure 11, for all tested positions, this value exceeds 20 dB. This is largely sufficient to support a single carrier modulation with 16 QAM. More precisely, for the worst case position (position number 3), we simulate the transmission of 1,500,000 random bits over a single carrier modulation transmission with a Raised Root Cosine (RRC) filter, 16 QAM, and a single tap receiver. For this simulation we chose an extreme value of Roll Off factor (0.001), to test the worst case scenario. We use the same simulation methodology detailed in [38], except that we use the current measured beamformed channel impulse response. The resulting measured bit error rate over 1,500,000 bits is zero. This means that the attainable BER, in this case, is estimated to be lower than $10^{-5}$.

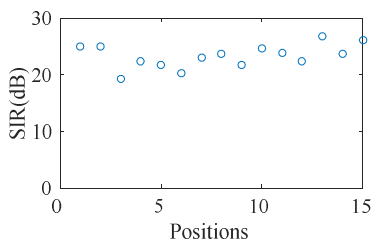

**FIGURE 11.** SIR with single carrier modulation.

This confirms that current standards for mobile networks have the potential to support single carrier modulations, with bandwidths as large as several MHz. Note that, by applying maximum ratio combining (MRC) at the receiver side (instead of MRT at the transmitter side), in an uplink transmission (instead of a downlink transmission) we would have obtained the same result: the channel after equalization would have been single tap. This means that after a receive matched filter, the channel is single tap and compatible with a single carrier modulation and a single-tap detector.

In the next sections, M-MIMO is used with much more antenna elements than in the current sub-section. It is therefore a reasonable assumption to consider that the beamformed or equalized channel (using transmit matched filtering or received matched filtering) will be single tap, i.e., not frequency selective. As a consequence, to derive the beamformed channel, one just needs to study the frequency-flat channel over the carrier frequency. Therefore, in the next sections, we will directly use a frequency-flat channel model for our performance studies and assume that the equalized or pre-equalized channel does not introduce interference between successive symbols in the time domain.

## IV. UPLINK TRANSMISSION: FIRST PERFORMANCE EVALUATION OF TRANSMIT SPATIAL MODULATION WITH REAL COMPACT RECONFIGURABLE ANTENNAS

In this section, for the first time, we present a complete performance evaluation of a wireless communication system using transmit spatial modulation with a real and existing compact reconfigurable antenna at the transmitter side and an M-MIMO antenna at the receiver side. The considered carrier frequency in this section is 2.45 GHz.

### A. MODELS OF REAL COMPACT RECONFIGURABLE ANTENNAS AND THE PROPAGATION CHANNEL

At the base station side, we consider a ULA of $M = 64$ elements spaced by half a wavelength as an example of M-MIMO antenna.

As for the object, two different real and existing compact reconfigurable antennas are considered and compared:

- A multiport switchable antenna [36] illustrated in Figure 12-a) that can generate $P = 4$ different radiation patterns, and that can therefore transmit $K = \log_2(P) = 2$ bits by using spatial modulation;
- A monoport reconfigurable antenna [40] illustrated in Figure 13-a) that can generate $P = 2$ different radiation patterns, and that can therefore transmit $K = \log_2(P) = 1$ bit by using spatial modulation.

The first one is half of the wavelength in size, whereas the second one only occupies one third. For both antennas, only one radiation pattern is activated at a time, either thanks to a RF switch connected to the multiport antenna or by commuting the diodes of the reconfigurable antenna [40].

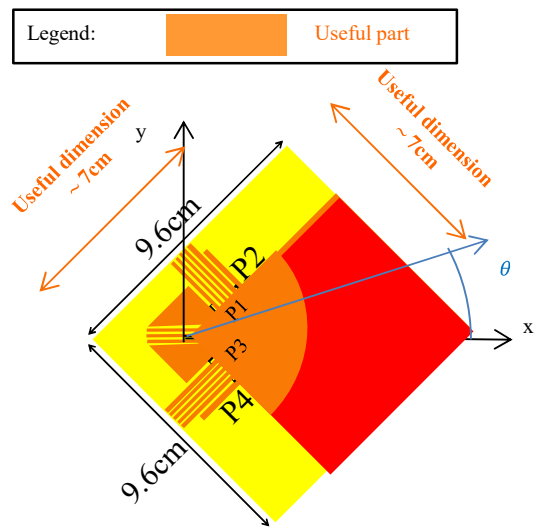

a) Antenna dimensions and convention for $\theta$

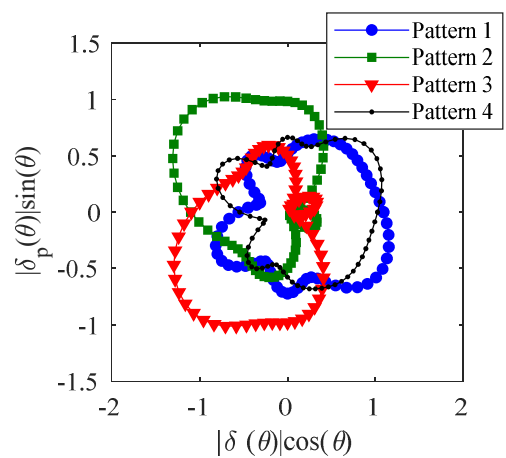

b) 2-D radiation patterns, where $\theta$ is the angle of arrival or departure and $\delta_p(\theta)$ is the antenna gain in the direction of $\theta$ for port $p$

FIGURE 12. Compact multiport switchable antenna model.

The two-dimensional (2D) propagation models, for the multiport switchable antenna and the monoport reconfigurable antenna are illustrated in Figure 14 and Figure 15, respectively. For the two compact reconfigurable antennas, the complex gain function $\delta_p(\theta)$ ($\delta_p: R \rightarrow C$) of each pattern $p$ in direction angle $\theta$ is numerically characterized by using Finite Difference Time Domain (FDTD) full wave simulation. The moduli of the radiation patterns are illustrated in Figure 12-b) and Figure 13-b), for the multiport switchable antenna and the monoport reconfigurable antenna, respectively.

The spatial correlation between two radiation patterns $p$ and $q$ of the same antenna is a key parameter in spatial modulation systems. The less correlated the patterns are, the more robust spatial modulation to noise is. The correlation is given by:

$$\psi^{p,q} = \frac{\left|\int_{\theta=0}^{2\pi} \delta_p(\theta)\delta_q^*(\theta)d\theta\right|^2}{\left(\int_{\theta=0}^{2\pi}|\delta_p(\theta)|^2 d\theta\right) \times \left(\int_{\theta=0}^{2\pi}|\delta_q(\theta)|^2 d\theta\right)}$$

Table I and Table II provide the correlations for the multiport switchable antenna and the monoport reconfigurable antenna, respectively.

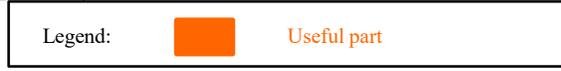
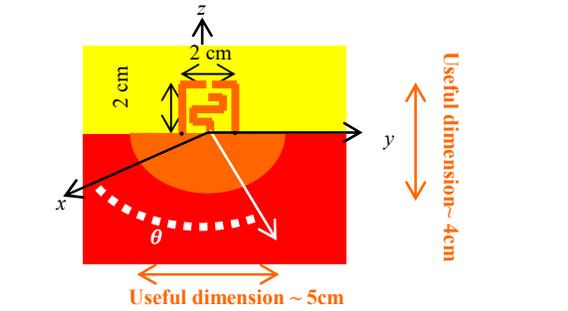

a) Antenna dimensions and convention for $\theta$

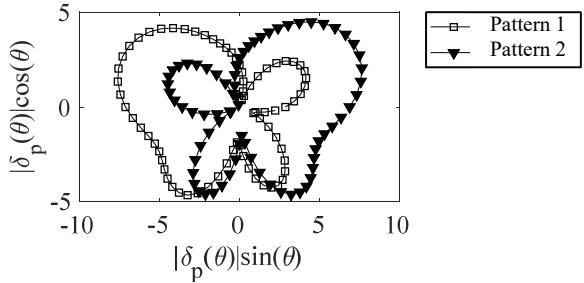

b) 2-D radiation patterns, where $\theta$ is the angle of arrival or departure and $\delta_p(\theta)$ is the antenna gain in the direction of $\theta$ for radiation pattern $p$

**FIGURE 13.** Compact monoport reconfigurable antenna model.

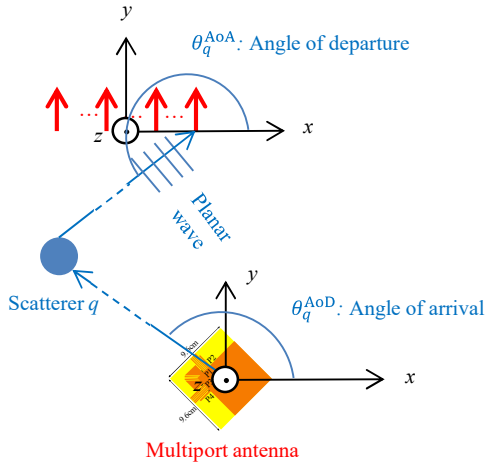

**FIGURE 14.** Propagation model for the multiport switchable antenna.

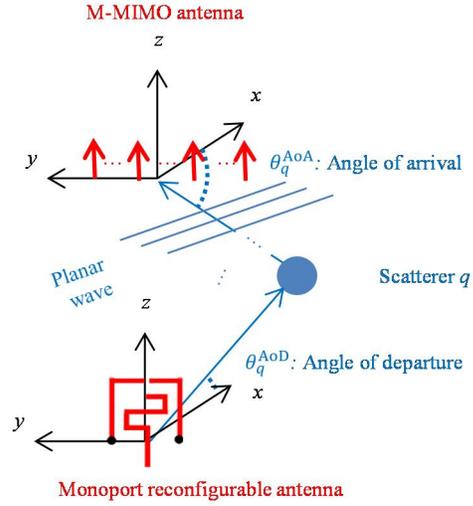

**FIGURE 15.** Propagation model for the monoport reconfigurable antenna.

TABLE I
CORRELATIONS $\psi^{p,q}$ BETWEEN ANTENNA PORTS P AND Q FOR THE MULTIPORT SWITCHABLE ANTENNA

| p\q | 1 | 2 | 3 | 4 |
|---|---|---|---|---|
| 1 | 1 | 0.0722 | 0.0593 | 0.0007 |
| 2 | 0.0772 | 1 | 0.0158 | 0.0522 |
| 3 | 0.0593 | 0.0158 | 1 | 0.0688 |
| 4 | 0.007 | 0.0522 | 0.0688 | 1 |

TABLE II
CORRELATIONS $\psi^{p,q}$ BETWEEN ANTENNA PORTS P AND Q FOR THE MONOPORT RECONFIGURABLE ANTENNA

| p\q | 1 | 2 |
|---|---|---|
| 1 | 1 | 0.3190 |
| 2 | 0.3190 | 1 |

We consider a single carrier modulation communication. As explained in section III, thanks to the M-MIMO antenna and assuming that a matched filter is used at the receiver, we can limit the study to a frequency-flat channel and assume that the channel does not introduce interference between successive symbols in the time domain. The wireless propagation channel between one antenna port of the transmitter and one antenna element of the receiver can be modeled by a complex gain.

We define $\mathbf{H} \in \mathbb{C}^{M \times P}$ as the channel between the object and the base station. More precisely, let $\mathbf{H}_{m,p}$ be the channel coefficient between the receive antenna $m \in [1,M]$ of the base station and the object when it is using the radiation pattern number $p \in [1,P]$. $\mathbf{H}_{m,p}$ includes both the wireless propagation and the radiation pattern $p$.

Regarding the model used for multi-path propagation, we consider a 2D wireless propagation model, with $Q$ random scatterers creating angular diversity in the channel. The elements of the M-MIMO antenna are spatially correlated. The path $q \in [1, Q]$ between the receive antenna $m$ of the base station and the connected object has a random complex gain $\Gamma_q \in \mathbb{C}$, a random angle of departure $\theta_q^{AoD} \in [0, 2\pi[$ and a random angle of arrival $\theta_q^{AoA} \in [0, 2\pi[$. $\Gamma_q$ is a Rayleigh fader with $E\left[\left|\Gamma_q\right|^2\right] = 1/Q$ and $\theta_q^{AoD}$ and $\theta_q^{AoA}$ are uniformly distributed over $[0, 2\pi[$. With these notations, the channel coefficient $\mathbf{H}_{m,p}$ is given by:

$$\mathbf{H}_{m,p} = \rho \sum_{q=1}^{Q} \Gamma_q \delta_p(\theta_q^{AoD}) e^{j\pi . \sin(\theta_q^{AoA})(m-1)}, \quad (1)$$

where $\rho$ is a normalizing factor. We choose $\rho$ such that:

$$\frac{\|\mathbf{H}\|^2}{PM} = 1.$$

In other terms, the average channel power per SISO antenna link is unitary.

*B SYSTEM MODEL*

For the compact multiport switchable antenna, we compare five different schemes to send a sequence $\mathbf{b}$ of $r$ bits, where, $\mathbf{b} = \mathbf{b}_1 \ldots \mathbf{b}_r \in \mathcal{B}^{(r^2)}$, and $r = 4$:
- "16QAM & Pattern $p = 1$";
- "16QAM & Pattern $p = 2$";
- "16QAM & Pattern $p = 3$";
- "16QAM & Pattern $p = 4$";
- "QPSK & SM4".

For the compact monoport reconfigurable antenna, we compare three different schemes to send a sequence $\mathbf{b}$ of $r$ bits, where $\mathbf{b} = \mathbf{b}_1 \ldots \mathbf{b}_r \in \mathcal{B}^{(r^2)}$, with $r = 4$:
- "8PSK & Pattern $p = 1$";
- "8PSK & Pattern $p = 2$";
- "QPSK & SM2".

For all schemes, for each symbol period, the object sends a sequence of $r$ bits $\mathbf{b} \in \mathcal{B}^{(r^2)}$ with a radiation pattern $p$. Among these $r > 0$ bits, $u > 0$ bits are sent using a complex modulation symbol $s \in S$, where $S$ is the pre-defined set of complex modulation symbols. $K \geq 0$ bits are sent using spatial modulation. So, the following holds:

$$r = u + K.$$

The definition of $S$, the values of $p$, $u$ and $K$ are scheme-specific and provided in Table III and Table IV for the schemes with the multiport switchable antenna and the monoport reconfigurable antenna, respectively. Note that, for "16QAM & Pattern $p$" and "8PSK & Pattern $p$", the radiation pattern $p$ is fixed, and spatial modulation is unused ($K = 0$). On the contrary, for "QPSK & SM4" and "QPSK & SM2" schemes, the pattern $p$ is variable and spatial modulation is used ($K > 0$).

TABLE III
SPECTRAL EFFICIENCY R FOR THE COMPACT MULTIPORT SWITCHABLE ANTENNA (WITH P=4 STATES), IN NUMBER OF BITS PER SYMBOL PERIOD

| Scheme | $p$ | $S$ | $u+K=r$ |
|---|---|---|---|
| 16 QAM with Pattern 1 | 1 | $S^{16QAM}$ | $4 + 0 = 4$ |
| 16 QAM with Pattern 2 | 2 | | |
| 16 QAM with Pattern 3 | 3 | | |
| 16 QAM with Pattern 4 | 4 | | |
| QPSM & SM4 | 1 to 4 | $S^{QPSK}$ | $2 + 2 = 4$ |

TABLE IV
SPECTRAL EFFICIENCY R FOR THE COMPACT MONOPORT RECONFIGURABLE ANTENNA (WITH P=2 STATES), IN NUMBER OF BITS PER SYMBOL PERIOD

| Scheme | $p$ | $S$ | $u+K=r$ |
|---|---|---|---|
| 16 QAM with Pattern 1 | 1 | $S^{16QAM}$ | $4 + 0 = 4$ |
| 16 QAM with Pattern 2 | 2 | | |
| 16 QAM with Pattern 3 | 3 | | |
| 16 QAM with Pattern 4 | 4 | | |
| QPSM & SM4 | 1 to 4 | $S^{QPSK}$ | $2 + 2 = 4$ |

More precisely, in the "QPSK & SM4" scheme, as illustrated in Figure 16, during each symbol period, $\mathbf{b}_1 \mathbf{b}_2$ is sent using QPSK. Simultaneously, $\mathbf{b}_3 \mathbf{b}_4$ is sent using the corresponding pattern number $p$ (based on Table V).

TABLE V
PATTERN-TO-BIT MAPPING RULE

| $\mathbf{b}_3 \mathbf{b}_4$ | 00 | 10 | 01 | 11 |
|---|---|---|---|---|
| $p$ | 1 | 2 | 3 | 4 |

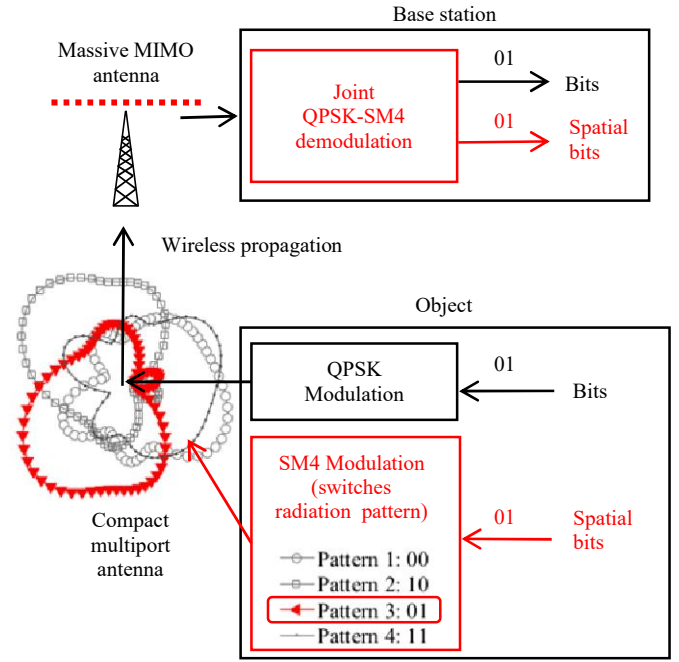

FIGURE 16. "QPSK & SM2" system model for the monoport compact antenna

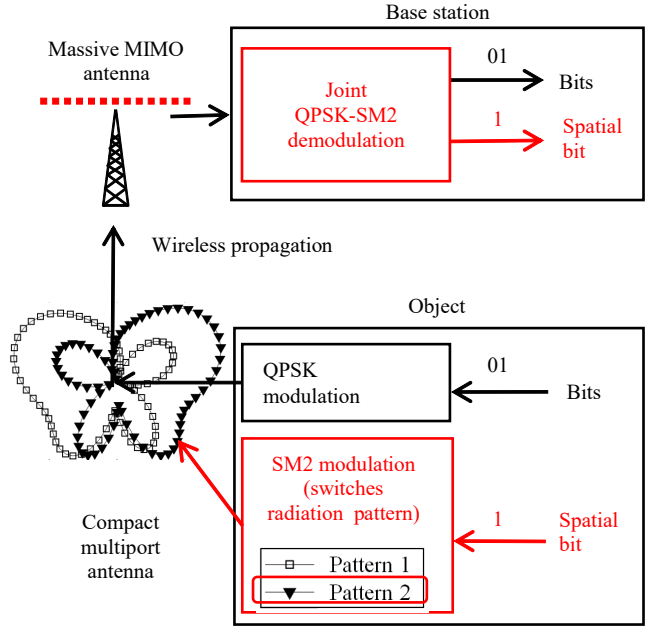

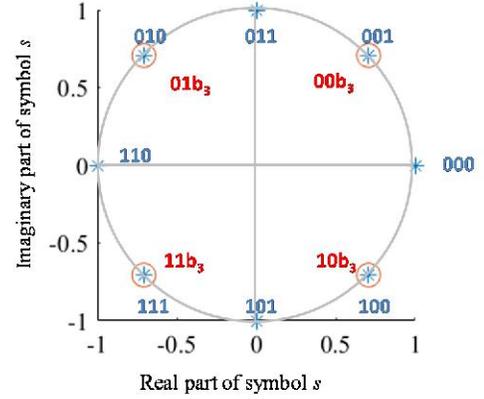

FIGURE 17. "QPSK & SM2" system model for the monoport compact antenna.

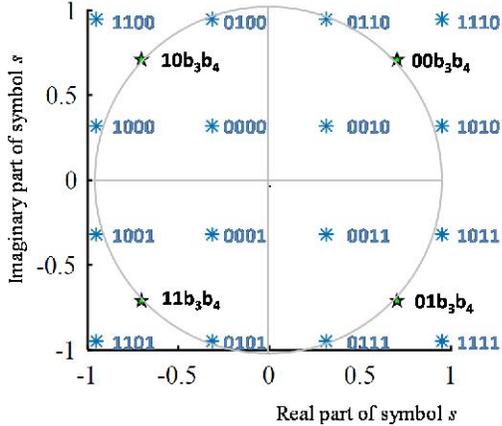

FIGURE 18. Constellations for 16QAM, "QPSK & SM4", with same average radiated power

In the "QPSK & SM2" scheme, as illustrated in Figure 17, during each symbol period, $b_1 b_2$ is sent using QPSK modulation. Simultaneously, $b_3$ is sent using the corresponding pattern number $p$ (pattern $p = 1$ corresponding to Bit "0" and $p = 2$ corresponding to Bit "1").

Figure 18 and Figure 19 illustrate the mapping between bits and symbols for the schemes with the multiport antenna and the reconfigurable antenna, respectively. Note that with the chosen definitions, $E[|s|^2] = 1$ for all schemes. In other terms, all schemes radiate the same transmit power per symbol, on average.

FIGURE 19. Constellations for 8PSK, "QPSK & SM2", with same average radiated power

For any scheme, we define $X$ as follows:
$X = \{\mathbf{x}^{(l)} \in \{0,1\}^{P \times 1} | 1 \leq l \leq r, \mathbf{x}_l^{(l)} = 1 \ \& \ \mathbf{x}_{p \neq l}^{(l)} = 0\}$.

Let $\mathbf{y} \in \mathbb{C}^{M \times 1}$ be the signal received over the $M$ elements of the M-MIMO antenna, $\mathbf{y}$ is given by:
$$\mathbf{y} = \mathbf{\Gamma}\left(\mathbf{H}\mathbf{x}s\sqrt{P_U} + \mathbf{v}\right),$$

where $\mathbf{x} \in X$, $s \in S$, $P_U$ is the transmit power, $\mathbf{v} \in \mathbb{C}^{M \times 1}$ is the vector of noise samples over the $M$ receiver chains of the base station, and $\mathbf{\Gamma} \in \mathbb{C}^{P \times M}$ accounts for the MRC receiver. More precisely, for $1 \leq p \leq P$ and $1 \leq n \leq M$, $\mathbf{\Gamma}_{p,n}$ is given by:
$$\mathbf{\Gamma}_{p,n} = \frac{\mathbf{H}_{n,p}^*}{\sum_{m=1}^{M}|\mathbf{H}_{p,m}|^2}.$$

We denote $P_{\text{NOISE}} = \frac{E[\|\mathbf{v}\|^2]}{M}$, as the average receiver noise power per antenna element at the base station side. We define an arbitrary signal to noise ratio metric $SNR$ that is common to all schemes:
$$SNR = \frac{P_U}{P_{\text{NOISE}}}.$$

We also assume that the receiver has a perfect estimate of $\mathbf{H}$ thanks to a previous training phase based on pilots. We assume that the receiver has computed and stored the following set of variables:

$$Y^{ref} = \{y = \mathbf{\Gamma H x} s\sqrt{P_u} \ |\mathbf{x} \in X \ \& \ s \in S\ \}.$$

Upon the reception of a new signal **y**, the receiver compares it to the variables of $Y^{ref}$ and determines the signal **ŷ** that minimizes the Mean Square Error:

$$\hat{\mathbf{y}} = \min_z \left\{ \frac{\|z-y\|^2}{M} | z \in Y^{ref} \right\}.$$

Then, the detected binary sequence $\hat{\mathbf{b}} = \hat{\mathbf{b}}_1 \dots \hat{\mathbf{b}}_r$ is deduced from **ŷ** by using the mapping rules illustrated in Figure 18 and Figure 19. The bit-error-rate (BER) can then be computed as follows:

$$BER = \frac{\sum_{k=1}^{r} |\hat{\mathbf{b}}_k - \mathbf{b}_k|}{r}.$$

We perform 250,000 simulation runs. For each simulation run:

- The parameter $SNR$ and the number of scatterers $Q = 10$ are fixed.
- We compute a random channel sample **H** based on randomly and independently generated parameters $\Gamma_q$, $\theta_q^{AoD}$, $\theta_q^{AoA}$, $\varphi_{m,q}^{AoD}$, $\Lambda_{m,q}$ and **v**.
- We generate a random sample of noise **v**.
- For each possible values of the sent sequence **b**, i.e. for all $\mathbf{b} \in \mathcal{B}^{(16)}$, we compute **y**, **ŷ** and **b̂** and the BER.

We average these BER values over all sent sequences, during a simulation run, and over all runs (i.e. over 1 million of bits). We then plot the result as a function of $SNR$ in Figure 20 and Figure 21, for the multiport antenna, and the monoport reconfigurable antenna, respectively.

### C. SIMULATION RESULTS AND CONCLUSIONS

Figure 20 shows that the compact monoport reconfigurable antenna with "QPSK & SM2" modulation is a solution to provide 3 bits/s/Hz, that is as efficient and as compact in size as another solution using a compact antenna of the same size with a conventional 8PSK modulation. However, the monoport reconfigurable antenna with "QPSK & SM2" is more energy efficient than 8PSK, as it advantageously uses constant envelop power amplifiers [41][42].

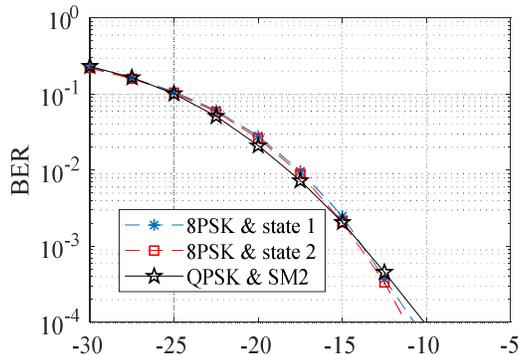

FIGURE 20. Performance with the monoport reconfigurable Antenna

Figure 21 shows that "QPSK & SM4" outperforms all other "16QAM & Pattern p" schemes. As for the previous antenna, "QPSK & SM4" is also more energy efficient than 16QAM, as it uses constant envelop amplifiers. Therefore, there is still room for improvement of the compactness of the antenna before the same BER vs SNR performance as 16QAM is attained.

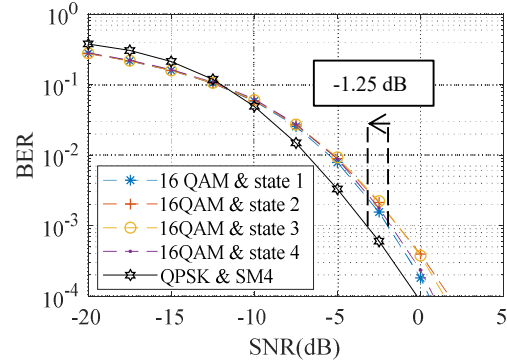

FIGURE 21. Performance with the multiport antenna

Based on these current results, as illustrated in Figure 22, one path for future improvements is to build a new monoport reconfigurable antenna, that is more compact than the current multiport antenna, and to deliver 4 bits/s/Hz with a constant envelop modulation (the same "QPSK & SM4" scheme for instance).

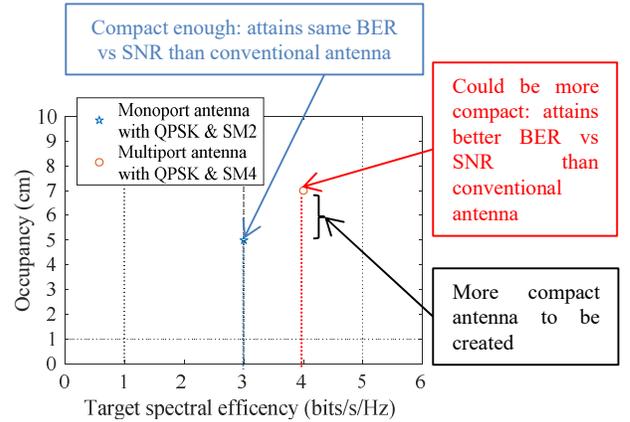

FIGURE 22. Path for improvement of the antenna for 4 bits/s/Hz

### V. DOWNLINK TRANSMISSION: FIRST PERFORMANCE EVALUATION OF TRANSMIT SPATIAL MODULATION WITH REAL COMPACT ANTENNAS

In this section, for the first time, we present a complete performance evaluation of a wireless communication system using receive spatial modulation with a real and existing compact antenna at the object side and an M-MIMO antenna at the network side. The considered carrier frequency in this section is 2.43 GHz.

## A. REAL COMPACT MULTIPORT ANTENNA

At the base station side, we consider a ULA of $M = 64$ elements spaced by half a wavelength as an example of M-MIMO antenna.

The same real and existing compact multiport antenna [36] as the one used in Section III, and illustrated in Figure 12 is used. However, this time, there is no switch, and all the ports are used simultaneously. A similar 2D channel model as the one used in section III-A is used. This time, we define $\mathbf{H} \in \mathbb{C}^{P \times M}$ as the channel from the base station to the object, including wireless propagation and antenna radiation patterns. More precisely, let $\mathbf{H}_{p,m}$ be the channel coefficient between the transmit antenna $m \in [1, M]$ of the base station and the antenna port $p \in [1,4]$ of the object. $\mathbf{H}_{p,m}$ includes both the wireless propagation and the radiation pattern of the port $p$.

As in Section III-A and as illustrated in Figure 23, we consider a 2D wireless multipath propagation model, with $R$ random scatterers creating angular diversity in the channel. The antenna elements of the M-MIMO are spatially correlated. The path number $q \in [1, Q]$ between the transmit antenna $m$ of the base station and the connected object has a random complex path gain $\Gamma_q \in \mathbb{C}$, a random angle of departure $\theta_q^{\text{AoD}} \in [0,2\pi]$ and a random angle of arrival $\theta_q^{\text{AoA}} \in [0,2\pi]$. $\Gamma_q$ is Rayleigh distributed with $E\left[\left|\Gamma_q\right|^2\right] = 1/Q$ and $\theta_q^{\text{AoD}}$ and $\theta_q^{\text{AoA}}$ are uniformly distributed over $[0,2\pi]$. With this notation, the channel coefficient $\mathbf{H}_{p,m}$ is given by:

$$\mathbf{H}_{p,m} = \rho \sum_{q=1}^{Q} \Gamma_q \delta_p(\theta_q^{\text{AoA}}) e^{j\pi.\sin(\theta_q^{\text{AoD}})(m-1)}, \quad (2)$$

where $\rho$ is a normalizing factor. We choose $\rho$ such that:

$$\frac{\|\mathbf{H}\|^2}{PM} = 1.$$

The average channel power per single antenna link is unitary.

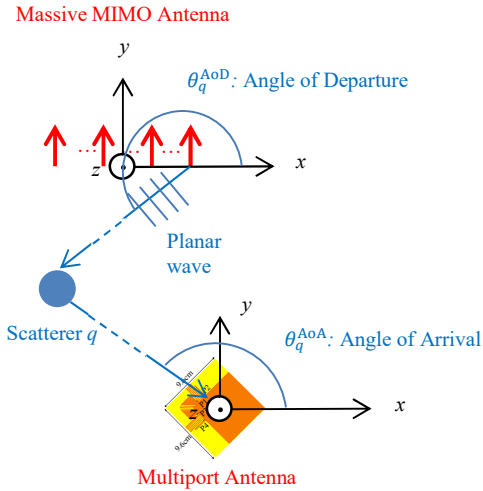

FIGURE 23. Propagation model for the multiport antenna

## B. SYSTEM MODEL

We compare the following five different schemes:
- "16QAM & Pattern $p = 1$";
- "16QAM & Pattern $p = 2$";
- "16QAM & Pattern $p = 3$";
- "16QAM & Pattern $p = 4$";
- "QPSK & SM4".

For all schemes, for each symbol period, the base station sends a sequence of 4 bits $\mathbf{b} = \mathbf{b}_1\mathbf{b}_2\mathbf{b}_3\mathbf{b}_4 \in \mathcal{B}^{(16)}$ using a complex modulation symbol $s$ and a precoder $\mathbf{\Gamma}^{(p)}$, with $p \in [1,4]$, picked among 4 stored precoders. The precoder $\mathbf{\Gamma}^{(p)} \in \mathbb{C}^{M \times P}$ is based on the MRT precoder and is defined as follows:

$$\mathbf{\Gamma}_{m,p}^{(p)} = \alpha^{(p)}(\mathbf{H}_{p,m})^*, \quad (3)$$

where, $\alpha^{(p)}$ is chosen so that:

$$\sum_{m=1}^{M}\left|\mathbf{\Gamma}_{m,p}^{(p)}\right|^2 = 1.$$

We denote by $S$ the set of complex modulation symbols. The definition of $S$ and the choice of $p$ are scheme-specific and provided hereafter for each scheme.

In "16QAM & Pattern $p$", the precoder $\mathbf{\Gamma}^{(p)}$ is fixed, and 16QAM modulation is used, hence:

$$S = S^{\text{16QAM}}.$$

The pattern-to-bit mapping in Table III is used.

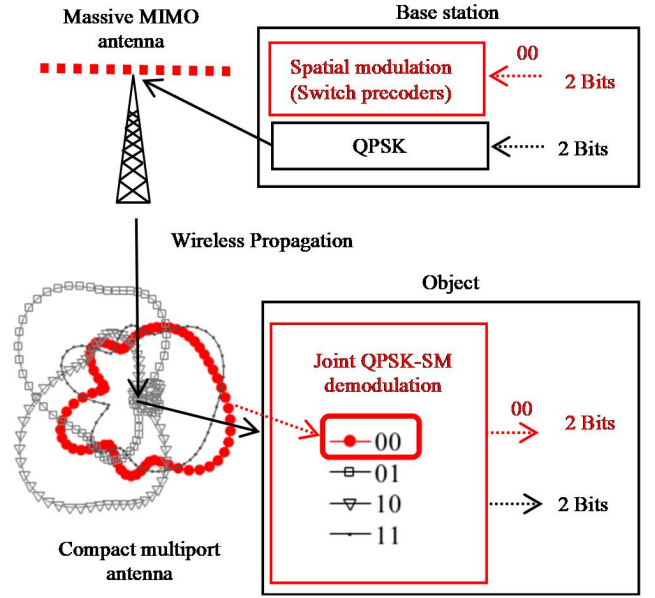

FIGURE 24. "QPSK & SM4" system model

In the "QPSK & SM4" scheme, as illustrated in Figure 24, during each symbol period, $\mathbf{b}_1\mathbf{b}_2$ is sent using QPSK modulation. Simultaneously, $\mathbf{b}_3\mathbf{b}_4$ is sent using the corresponding precoder $\mathbf{\Gamma}^{(p)}$ (using the pattern-to-bit mapping rule of Table III), hence:

$$S = S^{\text{QPSK}}.$$

The illustration of the mapping between bits and symbols in Figure 18 and the spectral efficiency given in Table III is still valid for the considered scheme. Note that with the chosen definitions, $E[|s|^2] = 1$ for all schemes. In other terms, all schemes require the same transmit power per symbol, in average.

We use the same notation $X$ as for section III:
$$X = \{\mathbf{x}^{(l)} \in \{0,1\}^{K\times 1} | 1 \leq l \leq 4, \mathbf{x}_l^{(l)} = 1 \;\&\; \mathbf{x}_{p\neq l}^{(l)} = 0\}.$$

Let $\mathbf{y} \in C^{M\times 1}$ be the signal received over the $P = 4$ ports of the object. With this notation, we obtain the following expression of $\mathbf{y}$:
$$\mathbf{y} = \mathbf{H}\mathbf{\Gamma}^{(p)}\mathbf{x}s\sqrt{P_u} + \mathbf{v},$$
where $\mathbf{x} \in X$, $s \in S$, $P_U$ is the transmit power and $\mathbf{v} \in \mathbb{C}^{M\times 1}$ is the vector of noise samples over the $P$ ports of the object. We denote $P_{\text{NOISE}} = E[\|\mathbf{v}\|^2]/P$, as the average receiver noise power at the object side per port. We define an arbitrary signal to noise ratio metric $SNR$ that is at least common to all schemes, as follows:
$$SNR = \frac{P_U}{P_{\text{NOISE}}}.$$

We assume that the receiver has perfect estimates of $\mathbf{H}$ and $\mathbf{\Gamma}^{(p)}$ thanks to a previous training phase based on pilots. We assume that the receiver has computed and stored the variables:
$$Y^{ref} = \{y = \mathbf{\Gamma}^{(p)}\mathbf{H}\mathbf{x}^{(p)}s\sqrt{P_u} \;|\; p \in [1,4] \;\&\; s \in S\}.$$

Upon the reception of a new signal $\mathbf{y}$, the receiver compares it to the set of variables $Y^{ref}$ and determines the signal $\hat{\mathbf{y}}$ that minimizes the Mean Square Error:
$$\hat{\mathbf{y}} = \min_z \left\{ \frac{\|z-y\|^2}{M} \;|\; z \in Y^{ref} \right\}.$$
Then, the detected binary sequence $\hat{\mathbf{b}} = \hat{\mathbf{b}}_1\hat{\mathbf{b}}_2\hat{\mathbf{b}}_3\hat{\mathbf{b}}_4$ is deduced from $\hat{\mathbf{y}}$ by using the mapping rule illustrated in Figure 18.

We perform a large number of simulation runs (100,000 runs) and use the same simulation methodology as in Section IV-B) to derive the average the BER versus $SNR$ curves of Figure 25.

### C. SIMULATION RESULTS

Figure 25 shows that "16QAM & SM4" outperforms all other schemes for the considered range of SNR and BER values. This is due to the combination of two effects. On the one hand, "16QAM & SM4" is less dense in the complex domain than the "16QAM & Pattern $p$" modulations, as illustrated in Figure 18. On the other hand, the patterns of the 4-Port compact antenna are low correlated, as shown by Table I.

We also observe that the BER performance of "QPSK & SM4" will be worse than that of 16 QAM modulation for higher SNR values. This is consistent with earlier studies on spatial modulation with conventional antenna arrays, led by at least two different independent teams in [43] (Fig. 5, Fig. 6 and Fig. 9) and in [44] (Fig. 9). These studies show that spatial modulation is outperformed by conventional modulations at high SNR.

To summarize the results, "QPSK & SM4" is a new modulation which, compared to 16QAM, is more energy efficient, in the considered ranges of SNR and BER values (larger than $10^{-4}$).

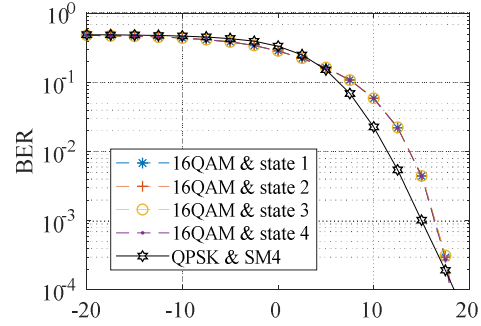

**FIGURE 25.** Simulation results

## VI. CONCLUSIONS AND FUTURE PERSPECTIVES

In this paper, for the first time, we have presented a complete performance evevaluation results of two new wireless communication systems for small connected objects, both based on spatial modulation and real existing compact antennas at the object side, and M-MIMO antennas at the base station side. Based on experimental results, we have shown that introducing M-MIMO in 5G networks enables the potential re-introduction of single carrier modulation based on spatial modulation. Our simulations take into account precise models of several real and existing compact antennas (obtained from actual prototypes designed and implemented for special application to single-carrier spatial modulation systems) at the object side: a compact monoport reconfigurable antenna with two states and a multiport switchable antenna with four states. We have compared the proposed spatial modulation systems with conventional modulations of the same spectral efficiency. Our results show that transmit spatial modulation with the monoport antenna attains the same BER versus SNR performance as 8PSK, but it allows one to use constant envelop amplifiers, which are less complex and more energy efficient. Transmit spatial modulation with the multiport antenna attains a better BER versus SNR than 16QAM, and still with constant envelop power amplifiers. For this higher spectral efficiency, a more compact antenna could still be designed. Our simulations also show that receive spatial modulation with the multiport antenna outperforms 16QAM. Future investigations will focus on the design of antennas providing optimum performance and compactness, for a given target spectral efficiency.